\documentstyle[epsf,12pt]{article}

\begin{document}

\begin{titlepage}

\parindent0em
\begin{flushright}
astro-ph/9508028
\end{flushright}
\begin{center}
{\LARGE\bf Triggering of Imaging Air Cherenkov Telescopes:
PMT trigger rates due to night-sky photons}

\vspace*{3cm}
{\large G.~Hermann, C.~K\"ohler, T.~Kutter, W.~Hofmann}

\vspace*{1cm}
{\small Max-Planck-Institut f\"ur Kernphysik, P.O.Box 103980,
69029 Heidelberg, Germany}

\vspace*{1cm}
\today

\vspace*{2cm}
\begin{abstract} Imaging air Cherenkov telescopes are usually
triggered on a coincidence of two or sometimes more pixels,
with discriminator thresholds in excess of 20 photoelectrons
applied for each pixel. These thresholds required to
suppress night-sky background are significantly
higher than expected on the basis of a Poisson
distribution in the number of night-sky photoelectrons
generated during the characteristic signal integration time.
We studied noise
trigger rates under controlled conditions using an artificial
background light source. Large tails in the PMT amplitude
response to single photoelectrons are identified as a
dominant contribution to noise triggers. The rate of such
events is very sensitive to PMT operating parameters.
\end{abstract}

\end{center}
\end{titlepage}


\newpage
\section{Introduction}

Imaging Air Cherenkov Telescopes (IACTs)~\cite{WeekesReview}
have evolved
into one of the most important tools for ground-based
$\gamma$-ray astronomy in the TeV energy range. In an IACT,
Cherenkov light emitted by particles of an extended air
shower is imaged by a large mirror onto a camera
consisting of an array of photomultiplier tubes (PMTs).
Mirror sizes range from a few to almost 100 m$^2$;
initially, cameras consisted of some tens of phototubes
(``pixels"), whereas the latest generation of cameras now
in operation or under construction frequently has
several hundred phototubes, resulting both in an enhanced
resolution of the image and in an improved field of
view. The orientation and shape of the Cherenkov image
are related to the direction and type of the primary
particle, and can be used to separate $\gamma$-rays from the
otherwise overwhelming background of charged cosmic
rays. The total light yield in the image provides a
measure for the energy of the primary.

To initiate the digitization of the signals from the
camera, and of the read out of data, a trigger signal is
required. Usually, this trigger signal is derived from
the camera itself.
In order to
suppress spurious signals, two or more camera pixels
have to fire in coincidence, with pulse heights
exceeding a threshold $q_\circ$. The threshold value
is usually quoted in units of the mean pulse height per
photoelectron.
The dominant source of
spurious triggers are photons of the night-sky
background (NSB) light, with a flux of about
$2 \cdot 10^{12}$ photons/m$^2$~sr~s
in the relevant spectral range from 300 to 600~nm
{}~\cite{night_sky,WiednerNSB,LorenzNSB,Greenwhich}.
For a telescope with 5~m$^2$ mirror area and an effective
pixel diameter of 0.4$^\circ$ this translates into a photoelectron
rate of about 30~MHz per pixel.
In order to prevent
random coincidences between pixels
from saturating the data acquisition system,
rather high trigger thresholds for the individual PMTs
have to be chosen. Typical telescopes~\cite{WhippleTelescope,
HEGRA_CT1} with a two-fold trigger
coincidence require trigger thresholds of
20 or more photoelectrons in each PMT. As we shall see,
these threshold values are generally much higher than expected based
on the Poisson fluctuation in the number of NSB photoelectrons.

Topic
of this paper is the study and interpretation of the single-PMT trigger rates
and
the optimization of the behaviour of the individual PMTs of a
camera. We will first
briefly review the expected night-sky trigger rates.
 We then describe the experimental setup
used to study the trigger rates of PMTs and present the
experimental results and their interpretation.

This work was carried out in preparation of the construction of
the HEGRA IACT system - an array of 5 Cherenkov
telescopes now under deployment at the Canarian island
of La Palma. These telescopes will have about 8.3 m$^2$
mirror area, 5 m focal length, and will be equipped with
cameras consisting of up to 271 PMT pixels with size
corresponding to $0.25^\circ$. It is planned to use Thorn-EMI 9083
PMTs, equipped with a short light funnel with a
hexagonal entrance area. An additional prototype telescope has
a mirror area of 5 m$^2$, and an initial camera
consisting of 37 Russian FEU-130 PMTs, each subtending a
field of view of $0.4^\circ$. A detailed description of this
telescope $-$ whose camera is meanwhile upgraded to 127
Thorn-EMI pixels $-$ is found in~\cite{HEGRA_CT1}.

\section{Expected single-PMT trigger rates and coincidence rates}

In the following we review
estimates for the rates at which individual pixels and the
trigger coincidence fire.

Let us assume, for simplicity, that the individual pixels of a camera
fire at identical rates $R_{pixel}$, and that all discriminator
output pulses have the same length $T$.
If the
coincidence unit reacts instantaneously, the rate $R_n$ of
random $n$-pixel coincidences is approximately (for
$R_{pixel} T \ll 1$):
\begin{equation}
R_n = n C_n R_{pixel}\left(R_{pixel}T\right)^{n-1}~~~.
\end{equation}
The coefficient $C_n$ gives the number of different $n$-pixel
combinations; for a two-fold coincidence of $M$ pixels, $C_n
= M(M-1)/2$. Typical coincidence windows $T$ range from 10 to 20~ns;
acceptable IACT trigger rates $R_{n}$ are usually in the range
of a few Hz. For a two-fold coincidence of the 37 pixels
of the HEGRA prototype telescope,
e.g., these numbers imply single-pixel trigger rates of
a few 100 Hz, many orders of magnitude below the 30-MHz rate at
which single photoelectrons are generated by NSB. Obviously,
trigger thresholds have to be chosen high enough such
that only rare fluctuations trigger a pixel.

To estimate the single-pixel rates $R_{pixel}$ for a given discriminator
threshold, a usual approach is the following:
the single-photoelectron
pulse response
of the PMT is approximated as a
rectangle of length $\tau$, the characteristic shaping time.
A discriminator with its threshold set at $q_\circ$
photoelectrons will then trigger if within a time $\tau$ at
least $q_\circ$ photoelectrons are generated. Assuming that the
individual night sky photons, and hence the
photoelectrons are uncorrelated, the rate $R_{pixel}(q_\circ)$
 can be estimated to
\begin{equation}
R_{pixel}(q_\circ) \approx R_{pe} {(R_{pe} \tau)^{q_\circ-1}
\over (q_\circ-1)!} e^{-R_{pe} \tau}
\label{eq1}
\end{equation}
where $R_{pe}$ is the rate at which photoelectrons are generated,
provided that the rate is small, $R_{pixel} \ll R_{pe}$.

In order to reduce random
IACT triggers, two factors appear to be re-le-vant: a
short coincidence resolution time $T$ is useful (but in
case of the usual 2-fold coincidence not too crucial,
given that the rate $R_2$ depends only linearly on $T$). More
important is a fast response $\tau$ of the PMT, ideally close
to the spread of a few ns of the genuine Cherenkov
photons; for the required large thresholds $q_\circ \gg 1$
this shaping time enters with a high
power.

A short estimate shows that based on these equations, a
threshold of about 7 photoelectrons should reduce the
random trigger rate of a single PMT of the HEGRA-CT1 to well
below 300 Hz, and hence the random coincidence rate $R_2$ to below one Hz.
Operational experience, on the other hand, shows that
much larger trigger thresholds are required, indicating
that there are additional sources of PMT pulse-height
fluctuations.

The rate estimates given above neglect a number of effects.
For example, the pulse-height discriminator is often AC-coupled
to the PMT. The threshold is hence applied relative to the
time-averaged amplitude, resulting in a small shift. Also,
the output pulse from the PMT is not rectangular. However,
it is easy to see that approximating the pulse by a rectangular
pulse of the same peak height and fwhm, pile-up is overestimated
rather than underestimated. Finally, the single-photoelectron pulse spectrum
generated by a PMT is usually not a narrow line, but instead a rather
broad spectrum of relative width $w \approx 0.5 - 1$. Hence, the rms
noise is increased by a factor $\sqrt{1+w^2} \approx 1.1 - 1.5$;
the required threshold $q_\circ$ should increase by a similar factor.
None of these factors explains the discrepancy between
estimated and actually required thresholds.

Electronics noise could be another source of random
triggers, but rate measurements with a closed camera lid usually
identify night-sky photons as the dominant source of random
triggers.

\begin{figure} [htb]
\begin{center}
\leavevmode
\mbox{
\epsfysize9.26cm
\epsfxsize13.5cm
\epsffile{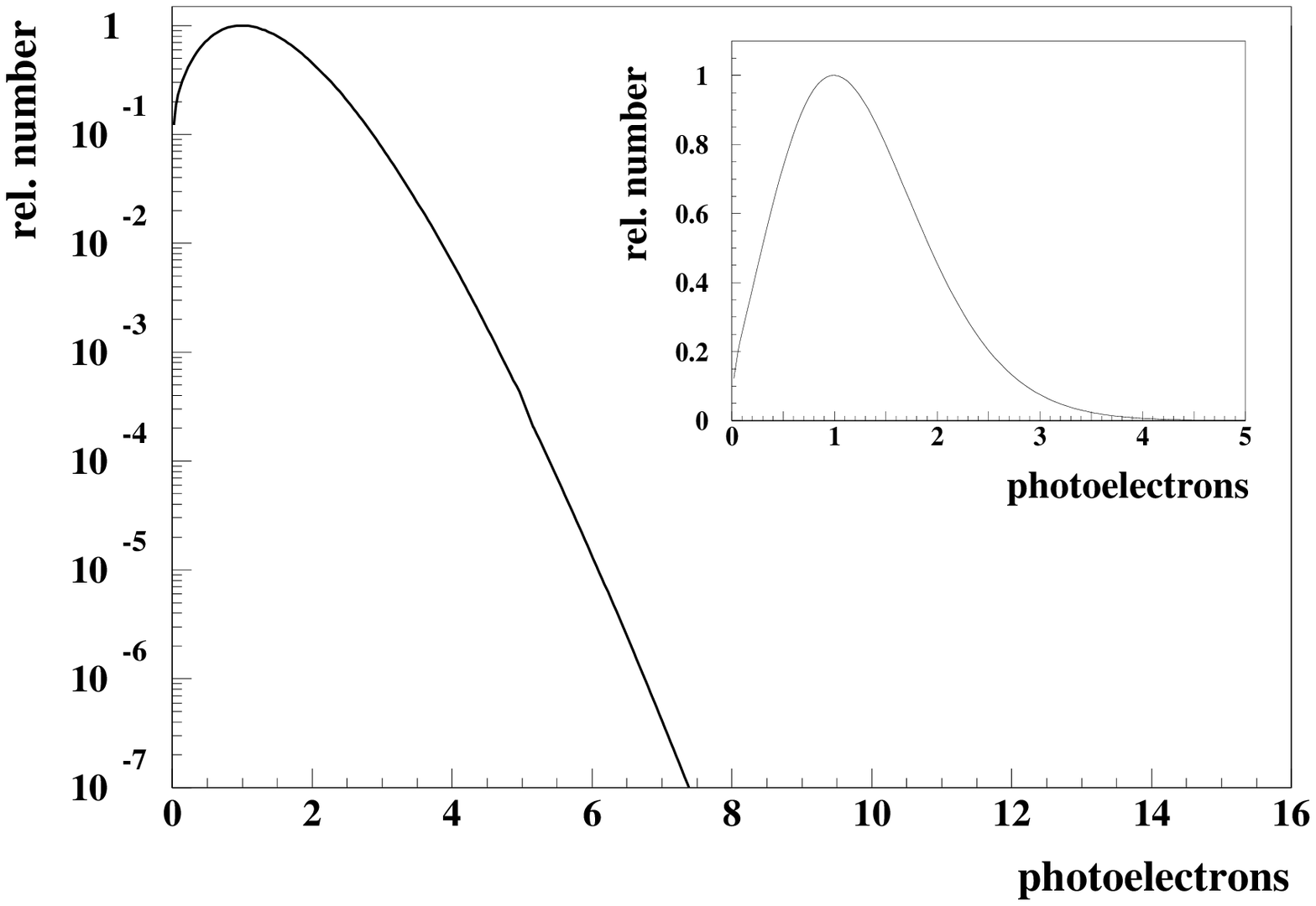}}
\caption[]{}
\end{center}
\label{fig_model_PE_dist}
\end{figure}

\begin{figure} [htb]
\begin{center}
\leavevmode
\mbox{
\epsfysize9.26cm
\epsfxsize13.5cm
\epsffile{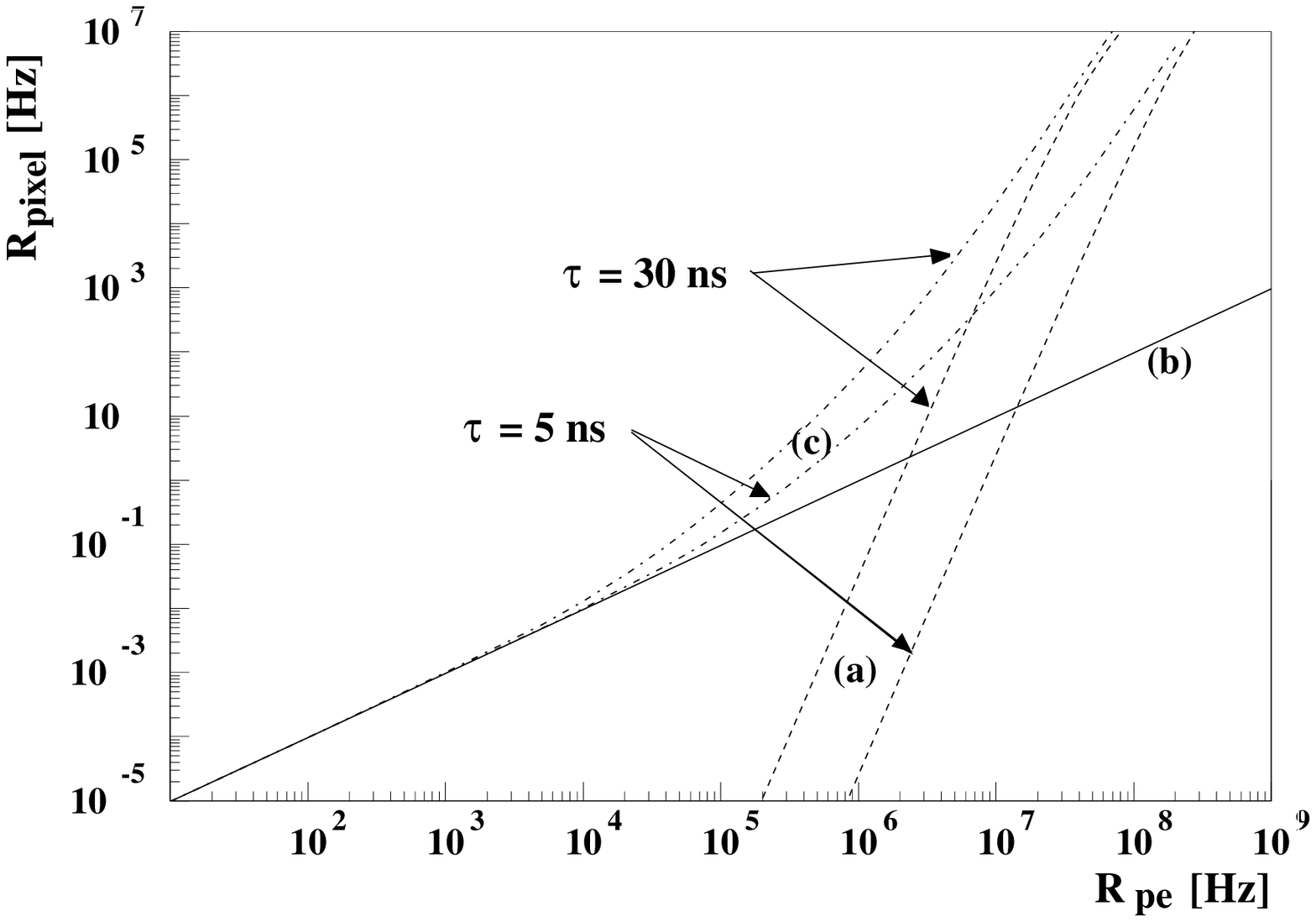}}
\caption[]{}
\end{center}
\label{fig_model_PE_rate}
\end{figure}

To provide more insight into the statistics of the amplification process,
we calculated the expected
distribution of pulse heights at the output of the PMT and
the trigger rates based on the assumption that
at each dynode the number of secondary electrons generated
by a primary electron follows a Poisson distribution.
Lacking an exact analytical expression, the response was
calculated by numerically convoluting the distributions for
the various stages of dynode amplification.
To describe the response of the PMT in terms of an
equivalent number of photoelectrons at its input,
the resulting pulse heights were rescaled to the
mean pulse height for a single photoelectron, corresponding to a
total gain $G = \Pi g_n$,
where $g_n$ is the gain of the n$^{\mbox{\footnotesize th}}$ dynode.
We used $g_1 = 5$ and $g_n = 2.5$ for $n > 1$,
corresponding roughly to the mean gain of BeCu dynodes at
the typical voltages applied in our tests~\cite{EMI_cat}.
Fig.~\ref{fig_model_PE_dist} shows the resulting distribution of
pulse heights for a single photoelectron.
Fig.~\ref{fig_model_PE_rate} illustrates the modeled trigger rate
as a function of the NSB photoelectron yield.
The modeled PMT trigger rate shows two distinct regimes:
\begin{itemize}
\item For a very small mean number of photoelectrons per integration
interval $\tau$, the trigger
probability is directly proportional to the NSB photoelectron rate.
In this regime, upward fluctuations
of the single-photoelectron response cause triggers. The rate is
much higher than expected on the basis of fluctuations in the
number of photoelectrons per integration interval, and
is quite insensitive to the pulse length $\tau$.
\item For higher photoelectron rates
the trigger probability grows with a higher power of the NSB photoelectron
rate.
In this regime, pile-up of photoelectrons causes triggers. The
trigger rate rises steeply with the pulse length $\tau$.
\end{itemize}

\section{Experimental setup and measurement procedure}

Rather than using actual night-sky light, the experimental
trigger studies were
carried out with an artificial source of background light, in order
to generate reproducible conditions. For all practical purposes,
only the rate at which ``night sky" photoelectrons are generated
should matter, and not details of the spectrum of the incident light.

In our setup, the PMT under test (usually
a Thorn-EMI 9083) is illuminated
by a dim DC light source; we use either a small
filament lamp or
a LED. In order to study the time structure of PMT pulses,
a pulsed fast LED as well as a laser with about 1~ns pulse length
is available.
The PMT output is coupled to a fast amplifier (LeCroy 612 AM),
which is connected via 15~m of RG178 cable to
a fan-out. A similar setup will be used in the real
telescopes, where a preamplifier is mounted in the camera, and the ADCs
and discriminators
are located in crates near the telescope. Despite its bandwidth limitations,
RG178 cable is used in order to limit the cross section and the weight of
the cable bundles. At the end of the cable, the PMT signal has a typical
width of about 5~ns (fwhm). One output of the fan-out is
routed to a discriminator (LeCroy 821),
the other, via a delay, to a charge-sensitive ADC (LeCroy 2249A) gated by the
discriminator signal. Alternatively, a Flash-ADC system
is used to record the pulse shapes, in particular to search
for afterpulses. Discriminator trigger rates are measured with a scaler.
The PMT current is monitored by switching the PMT output to an electrometer.
During a sequence of measurements, either the discriminator
threshold or the light intensity is varied. To investigate the
influence of the signal shaping time,
shaping amplifiers (Canberra 2111) could be added,
resulting in pulse lengths of 15~ns or 30~ns fwhm.

\begin{figure} [htb]
\begin{center}
\leavevmode
\mbox{
\epsfysize6.7cm
\epsfxsize13.5cm
\epsffile{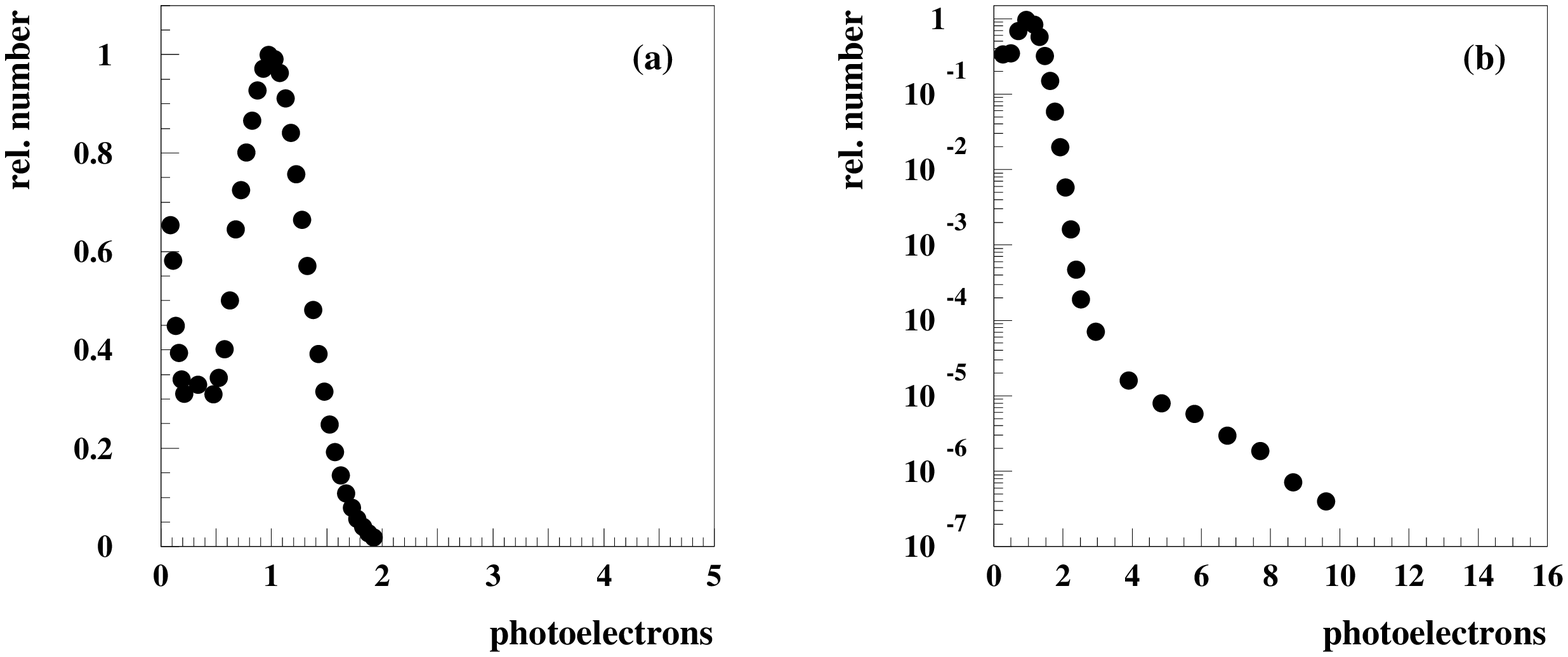}}
\caption[]{a,b}
\end{center}
\label{fig_pe_peak}
\end{figure}

Two critical constants have to be determined in order to calibrate the
system: the relation between the discriminator threshold voltage and the
equivalent number of photoelectrons, and the relation between the
PMT DC current and the (otherwise uncalibrated) photon flux. We use the
single-photoelectron peak to determine these coefficients. The charge
spectrum measured at low light levels and with a low threshold setting
is shown in Fig.~\ref{fig_pe_peak}(a),
already corrected for the ADC pedestal
and after subtraction of the dark-count spectrum accumulated over the same
time~\footnote{We note that the spectrum shows a
feature not reproduced in the
simulation: an excess at low pulse heights, which fills up the region
below the peak and which generates a peaking near zero charge. The origin
of these events is quasielastic backscattering of the photoelectron
from the first dynode, without multiplication~\cite{PM_handbook}.
It can also happen that
the photoelectron misses the first dynode entirely, and lands directly
on the second dynode. In  either case, the overall gain is reduced
by the gain of the first dynode, and because of the smaller gain of
the second dynode, the peak is smeared out.}.
With a higher discriminator
threshold setting, a relatively sharp cutoff in the spectrum is seen. Adjusting
the threshold until the cutoff is at the position of the
single-photoelectron peak
allows a rather precise calibration of the threshold in units of
photoelectrons, after correction for offsets and pedestals.
At lower operating voltages, the single-photoelectron peak
is not always well enough separated to use this technique.
In such cases, the calibration factor is determined at a higher
voltage and then scaled down in proportion to the PMT gain
as derived from the ratio of the DC currents at the two voltages,
measured at the same light flux.
For an asymmetric single-photoelectron peak such as the one seen in
Fig.~\ref{fig_pe_peak}, a principal
question is whether the calibration should refer to the peak position,
i.e., the {\em most probable} pulse height, or to the {\em average}
pulse height. We use the peak position as the easier and more reliable
measurement. We estimate the systematic
difference between the most probable and the average pulse height to be
less than 10 to 15 \%.

The single-photoelectron rate $R_{pe}$
 for a given light intensity can be obtained
either by directly counting events with a low threshold, or based on
the PMT average current $I$ and the (average) charge gain $G$ of the PMT,
\begin{equation}
R_{pe} = {I \over Ge}~~~,
\end{equation}
with the elementary charge $e$. We determine the charge gain using
the position of the single-photoelectron peak in the ADC spectrum.
With either method to measure $R_{pe}$, the behavior of the spectrum at very
low pulse heights causes systematic uncertainties.
The agreement between the two methods to
determine the photoelectron rate $-$ counting or current measurement $-$
depended somewhat
on the particular PMT used. Typically, the values agreed within 10 to 15\%.

In summary, we estimate that scale uncertainties in the calibration  of the
discriminator
threshold in units of photoelectrons amount to 15 \%, and that the
photoelectron rate has a similar systematic uncertainty.

\clearpage

\section{PMT trigger rates}

\begin{figure} [htb]
\begin{center}
\leavevmode
\mbox{
\epsfysize13.5cm
\epsfxsize13.5cm
\epsffile{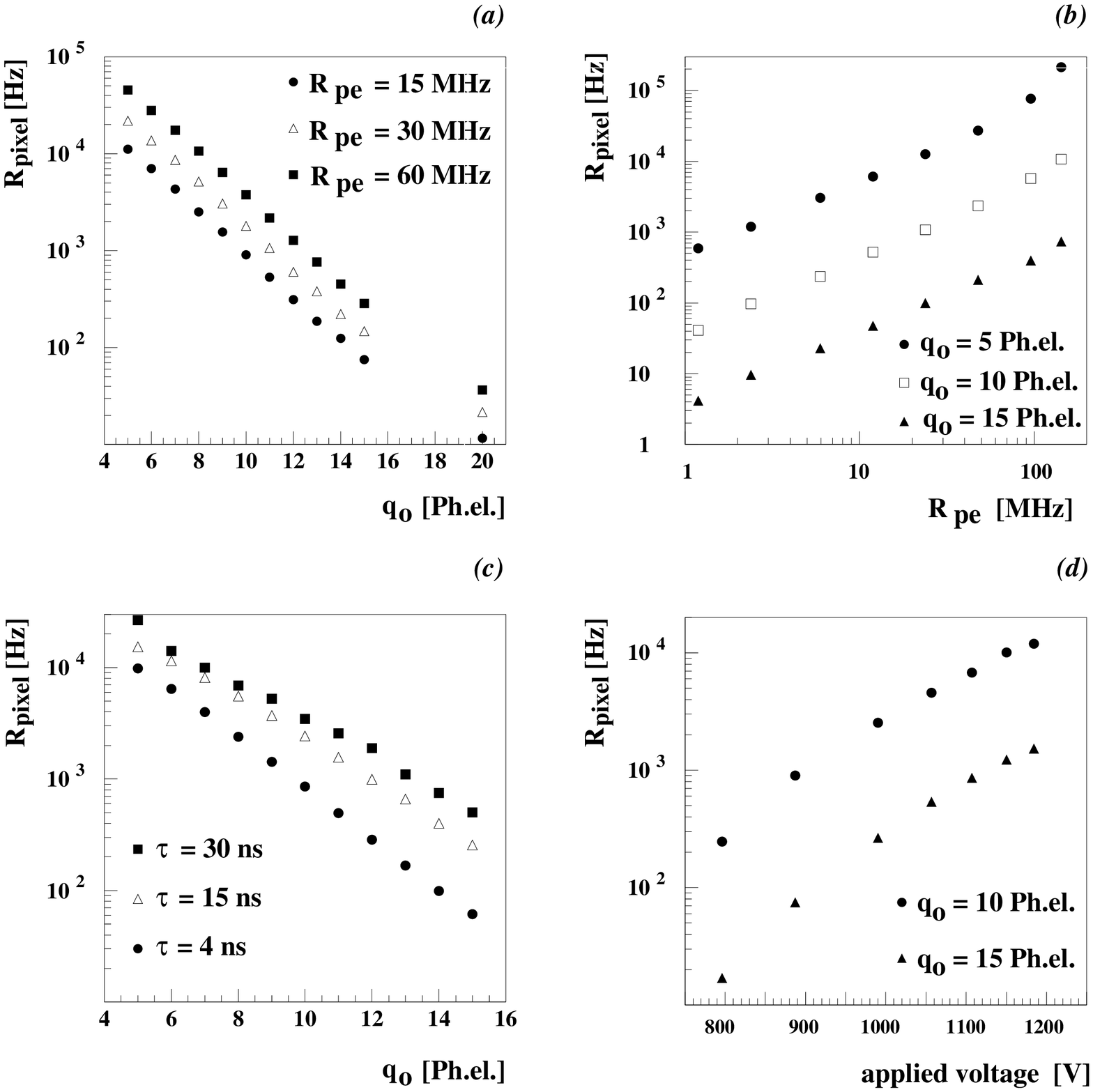}}
\caption[]{a-d}
\end{center}
\label{fig_results}
\end{figure}

Fig.~\ref{fig_results}
shows a typical set of measurements carried out using an
Thorn-EMI 9083 PMT. Operated as 10-dynode PMT, the tube provides
too much gain for use in an IACT, resulting in large anode
currents for typical NSB levels. Therefore, the signals were
derived from the 9th dynode, with the 10th dynode and the anode
connected to the same potential. We used a resistive voltage
divider, with resistor ratios 2:1.5:1:1... (starting from the
photo cathode). The divider was optimized to provide maximum
homogeneity and size of the active surface, and good linearity.

In Fig.~\ref{fig_results} we illustrate how the PMT trigger
rates depend on various parameters, such as the trigger
threshold $q_\circ$ (a), the light level (expressed in terms of
the photoelectron rate) (b), the pulse length (c) and the
high voltage applied to the PMT, with the trigger threshold
kept constant in units of photoelectrons (d).
Statistical errors
on the data points are generally negligible, except for
low rates in the Hz region; even there, statistical errors are
generally below 10\%.
Complete sets of measurements were repeated after
several days, including the determination of
calibration constants, and showed that rates are
reproduced within 15\%. For a given PMT current, illumination
with a filament lamp and a LED results in identical trigger rates,
within 5\%. Different
PMTs of the same type (and from the same production batch) showed
similar trigger rates, within a factor of 2.

The results shown in Fig.~\ref{fig_results} (a),(b),(c)
agree qualitatively with our
expectations; trigger rates drop with increasing threshold, with decreasing
light level, and with decreasing shaping time.
Quantitatively, they are however in striking contrast with the
model calculations (Fig.~\ref{fig_model_PE_rate}):
\begin{itemize}
\item With increasing threshold, trigger rates fall much more slowly than
expected on the basis of a Poisson distribution in the number of
photoelectrons, even including the gain fluctuations.
\item The absolute trigger rates for large thresholds are several orders
of magnitude larger than expected.
\item The trigger rate depends linearly on the light yield over
almost the entire range of light levels. Only for the highest levels
the dependence steepens.
\item The influence of the pulse length is relatively modest; based
on eq.~\ref{eq1} a change from 5 ns pulse length to 30 ns at a threshold
$q_\circ$ = 15 should increase the rate by about 8 orders of magnitude,
compared to an observed increase of less than 10.
\item Finally, as the single-photoelectron peak narrows with increasing
high voltage, one would expect the rate to decrease slightly, when the
voltage is increased, and the threshold is adjusted to correspond to
a fixed number of photoelectrons. In contrast, the data show a
significant increase of the rate with the PMT operating voltage.
For $q_\circ$ = 15, e.~g., the trigger rate rises with the
1.8$^{\mbox{\footnotesize th}}$~power
of the PMT gain.
\end{itemize}
The obvious interpretation of these observations is that gain fluctuations
have a much larger tail towards large pulse heights than expected; these
gain fluctuations dominate over the Poisson fluctuations in the number of
photoelectrons over nearly the entire range in $R_{pe}$. The presence of such a
tail
in the PMT response is indeed evident from the charge spectrum
of single-photoelectron signals, once it is plotted on a logarihmic
scale (Fig.~\ref{fig_pe_peak} (b)). At levels below about $3 \cdot 10^{-3}$
of the
peak, a long and flat tail takes over from the steep descent of charge
spectra right beyond the peak. In the region of IACT trigger thresholds,
this tail dominates. The tail is
not caused by background events unrelated to the illumination; the fraction of
events in the tail is constant over a wide range of light levels.
The observation of this tail
is not limited to the Thorn-EMI 9083 PMT; similar behaviour was observed
for the Philips XP2020, and for the (in its construction completely
different) Hamamatsu R5600P, although the relative level of the tail
differed.

\section{Origin of the amplitude tail and
optimization of PMT trigger properties}

\begin{figure} [htb]
\begin{center}
\leavevmode
\mbox{
\epsfysize18.9cm
\epsfxsize13.5cm
\epsffile{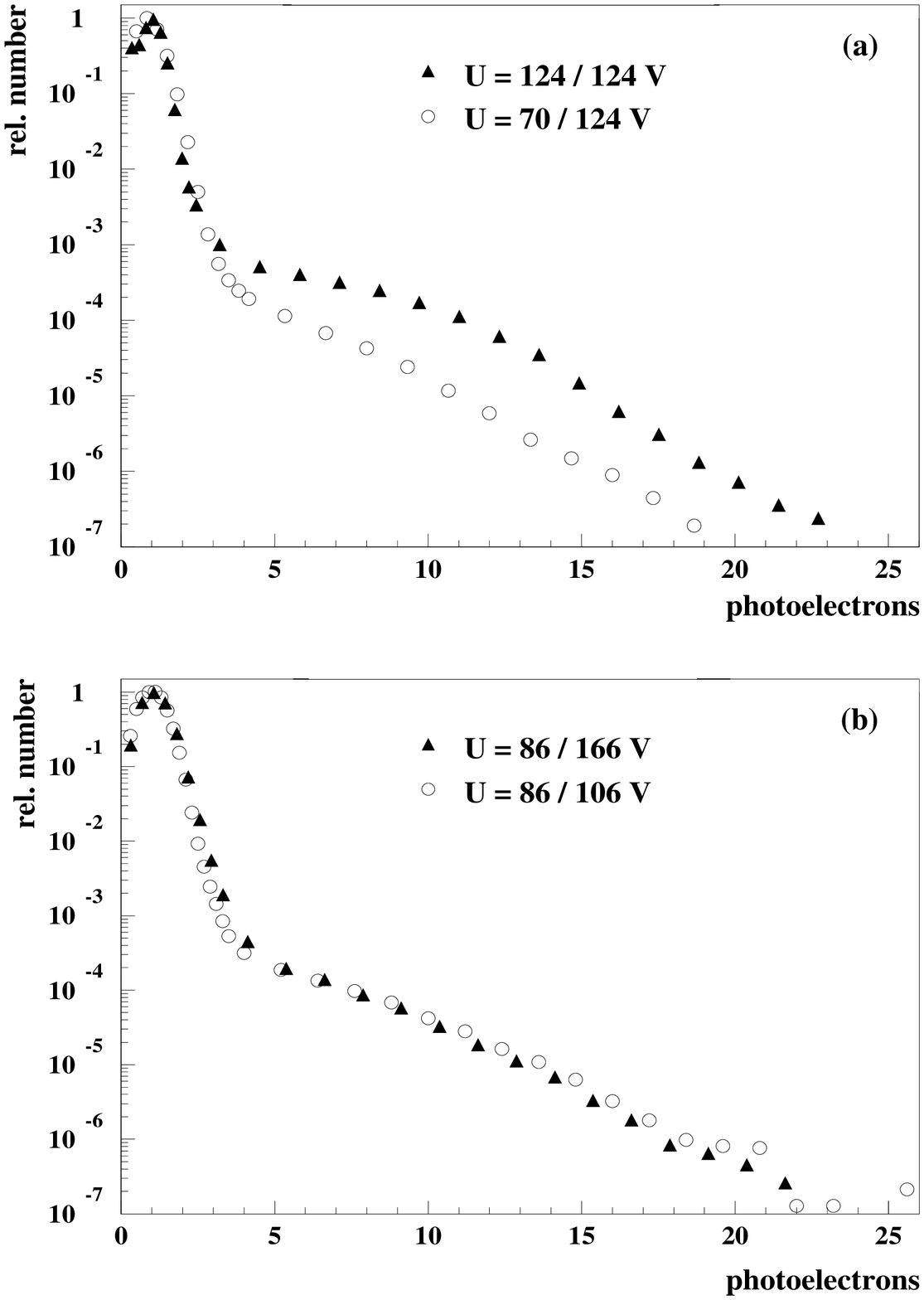}}
\caption[]{a,b}
\end{center}
\label{fig_HV_dep}
\end{figure}

\begin{figure} [htb]
\begin{center}
\leavevmode
\mbox{
\epsfysize9.26cm
\epsfxsize13.5cm
\epsffile{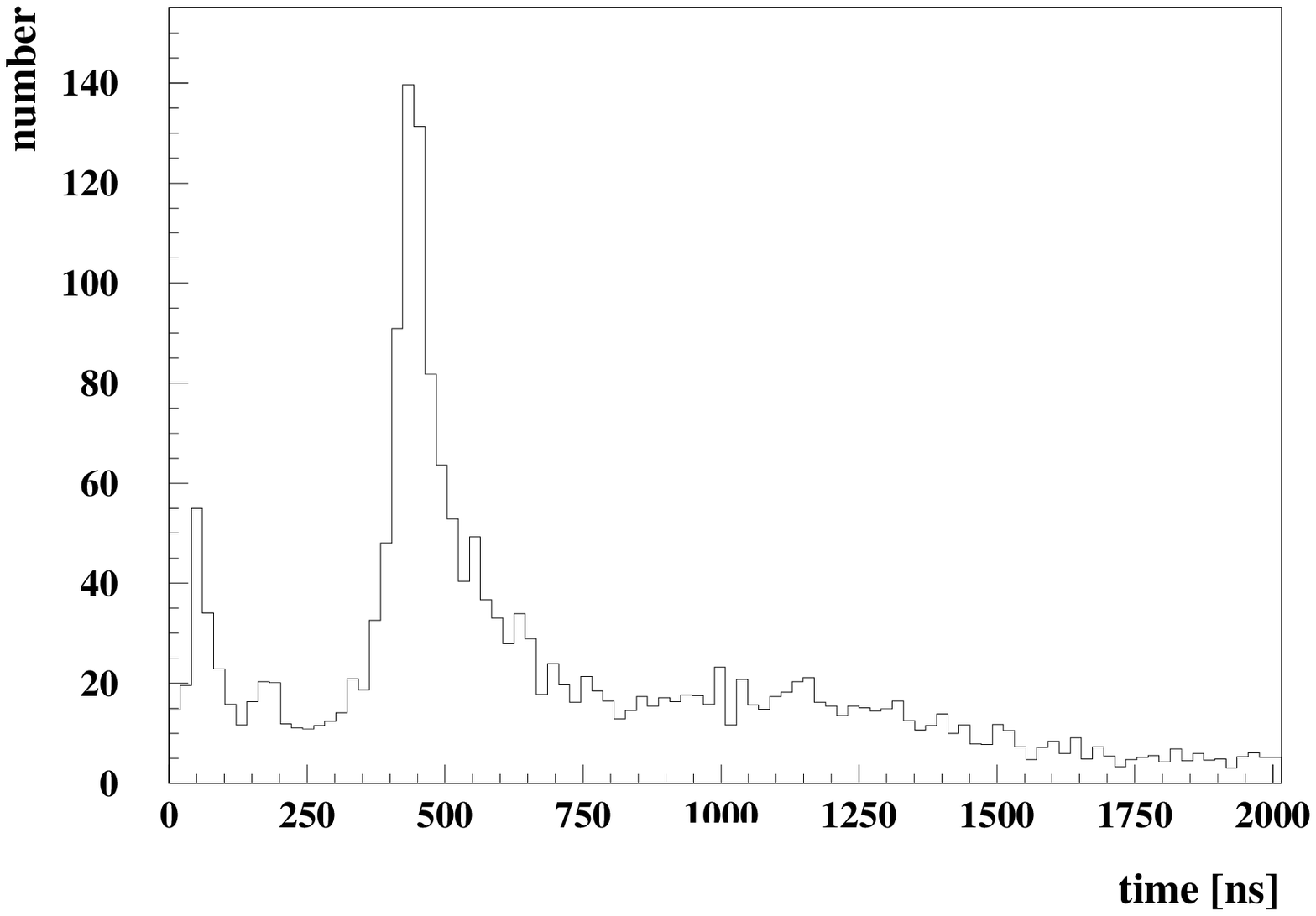}}
\caption[]{}
\end{center}
\label{fig_time_dep}
\end{figure}

\begin{figure} [htb]
\begin{center}
\leavevmode
\mbox{
\epsfysize9.26cm
\epsfxsize13.5cm
\epsffile{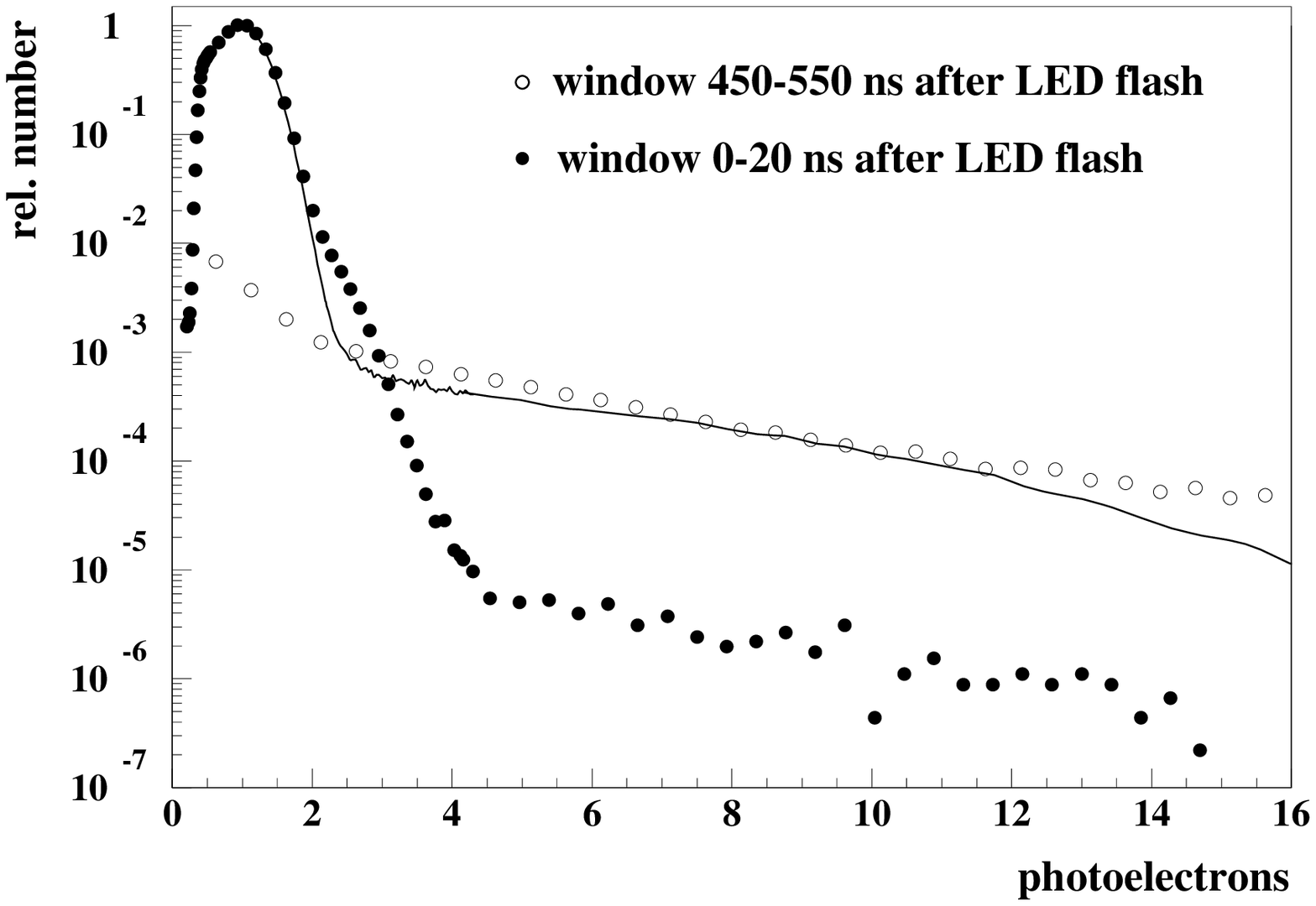}}
\caption[]{}
\end{center}
\label{fig_early_late}
\end{figure}

In order to identify the origin of the large pulses, which cause
the high trigger rates, we considered first the high-voltage dependence of
the phenomenon. The voltage of the early stages (photocathode
to 4th dynode) and of the later stages were varied independently
(see Fig.~\ref{fig_HV_dep}). The level of the tail is sensitive
to the field strenght in the early stages, and quite insensitive
to the later stages. As a candidate mechanism to generate the
large pulses, the ionization of
rest gas on or near the first dynode emerged. Studies with the
fast LED as a pulsed light source and the
Flash-ADC system to record the time history of the signals
did indeed reveal afterpulses following about 500~ns after
the light pulse (Fig.~\ref{fig_time_dep}). This delay is
consistent with the propagation time~\cite{PM_handbook47}
of an ion back to the photocathode,
where its impact releases a large number of electrons, which are
then subsequently amplified and generate a large output pulse.
Varying the voltage $V$ applied to the PMT, the delay $\Delta T$ was
observed to vary as
expected for such a mechanism, $\Delta T \propto 1/\sqrt V$.
Using the pulsed light source, we could separately record
the pulse height distributions of both prompt and delayed
pulses. Fig.~\ref{fig_early_late} illustrates that indeed
for prompt pulses (first 20~ns after the LED pulse) the
large-amplitude tail is significantly reduced, whereas the delayed pulses
(around 450 to 550~ns)
include the large-pulseheight tail causing most of the triggers
for large $q_\circ$.

What do these observations imply for the optimization of IACT
trigger schemes? Given the high photoelectron rate, an active
veto of delayed pulses seems excluded, since it would generate
enormous dead time. Since a major source of rest gas in PMTs
is the diffusion of Helium in the atmosphere through the PMT
walls~\cite{PM_handbook47}, an effective reduction of the rest gas
pressure appears difficult. A relatively simple means, however,
is the reduction of the voltage applied to the PMT; a lower
field in the space between photo cathode and 1st dynode reduces
both the probability for ionization, and the reacceleration of
ions. Usually, PMT voltages are optimized for optimal electron
collection and a good single-photoelectron peak. Given that at least
some telescopes operate with conversion factors of about 1 ADC channel
per photoelectron, a slightly-wider than optimal single-photoelectron
peak seems tolerable. We conclude that the voltage between
photocathode and 1st dynode should be chosen near the minimum
required for homogenous electron collection and response.

\subsection*{Acknowledgements}

We thank T. Lohse for contributions to the numerical modeling of the
PMT trigger rates. We acknowledge discussions with members of the HEGRA-IACT
group, in particular with R.~Mirzoyan. After completion of this work,
we learned that he arrived at similar conclusions concerning IACT trigger
rates.

\clearpage

\section*{Figure captions}
\parindent0em

{\bf Figure 1.}{ Distribution of PMT output pulse heights, scaled to the mean
output for one photoelectron, modeled assuming a Poisson distribution
in the number of secondary electrons, with gain 5 for the
first dynode, and gain 2.5 for the others. The Distribution is shown
with logarithmic and linear (small picture) scale for the relative
number of events.}

\vspace*{0.7cm}
{\bf Figure 2.}{ Modeled PMT trigger rate as a function of the NSB
photoelectron
yield, for 5~ns and 30~ns PMT output pulse width, and for a trigger
treshold $q_\circ=5$. The curves illustrate separately the effects of
poissonean
fluctuations in the arrival of photoelectrons (a), as estimated by
eq.~\ref{eq1},
and of fluctuations
in the amplification of single photoelectrons (b), as well as the full
numerical calculation including both effects (c).}

\vspace*{0.7cm}
{\bf Figure 3.}{ (a) Typical measured single-photoelectron spectrum of an
Thorn-EMI 9083 PMT,
after subtraction of the dark-count rate. The PMT is illuminated with
DC light, and the readout is triggered by the PMT signal.\newline
(b) Same, but with logarithmic scale.}

\vspace*{0.7cm}
{\bf Figure 4.}{ Trigger rates measured using a Thorn-EMI 9083 PMT, with the
signal derived from the 9th dynode. The nominal operating voltage
was 887~V.
(a) Trigger rate as a function of the discriminator threshold
$q_\circ$, for different light levels (expressed in terms of the
photoelectron rate).
(b) Trigger rate as function of light level, for different
discriminator thresholds $q_\circ$.\newline
(c) Trigger rate as a function of the discriminator threshold,
for different shaping time constants, resulting in 5, 15, and 30 ns
pulse length (fwhm).
(d) Trigger rate as a function of the high voltage applied
to the PMT, with discriminator threshold adjusted to correspond
to a fixed number of photoelectrons, $q_\circ=10$ and $q_\circ=15$.}

\vspace*{0.7cm}
{\bf Figure 5.}{ Charge spectrum obtained for low-level (single-photoelectron)
illumination of the PMT, with a resistive voltage divider 2R-1.5R-1R-1R-1R...
(starting from the photocathode). (a) For two different voltage levels
in the early part of the chain, up to dynode 4, and (b) for two different
voltage levels in the late part, dynodes 4 to 9. The voltage levels given
refer to the voltage across a resistor 1R, in the early and the late part
of the dynode chain.}

\vspace*{0.7cm}
{\bf Figure 6.}{ Arrival time $t$ of charge pulses from the PMT, following the
illumination
with a light pulse at the single-photoelecton level at $t\approx0$. The
output of the PMT was recorded by a Flash-ADC system. The prompt
pulses near $t \approx 0$ are suppressed.}

\vspace*{0.7cm}
{\bf Figure 7.}{ Charge spectrum of ``prompt" (20~ns window after the LED
flash)
and ``delayed" (450-550 ns after the LED flash) PMT pulses, for
single-photoelectron illumination with a pulsed LED. The solid line represents
the spectrum obtained if no correlation to the LED flash is requested.}

\newpage

\vspace*{3cm}

\end{document}